\documentstyle[pra,aps]{revtex}
\begin{document}
\draft
\title{Extracting an arbitrary relative phase from a multiqubit two-component
entangled state}
\author{Chui-Ping Yang and Siyuan Han}
\address{Department of Physics and Astronomy, University of Kansas, Lawrence,\\
Kansas 66045}
\maketitle

\begin{abstract}
We show that an arbitrary relative phase can be extracted from a multiqubit
two-component (MTC) entangled state by local Hadamard transformations and
measurements along a single basis only. In addition, how to distinguish a
MTC entangled state with an arbitrary entanglement degree and relative phase
from a class of multiqubit mixed states is discussed.
\end{abstract}

\pacs{PACS number{s}: 03.67.Lx, 03.65.Ud, 85.25.Dq}
\date{\today }

%\narrowtext

Entanglement plays an important role in quantum information processing and
communication, such as quantum computation, secret sharing, teleportation,
and quantum key distribution [1-3]. Over the last decade, characterizing
or/and quantifying entanglement has been recognized as one of central tasks
in quantum information science. Correspondingly, there have been many
efforts in this direction, for both pure and mixed states as well as for
bipartite and multipartite systems. These include entanglement of formation
[4], entanglement of distillation [5], relative entropy of entanglement [6],
negativity [7], geometric measure of entanglement [8], and so on.

On the other hand, there is a strong current interest in the detection of
quantum entanglement. To verify whether a qubit system is indeed prepared in
a desired entangled state, many methods for the entanglement detection have
been presented. Among them, there are proposals based on the structural
approximations [9], the semidefinite program [10], and the entanglement
witnesses [11-14]. In addition, the idea of inferring quantum entanglement
using a single measurement basis has been proposed by Schelpe {\it et. al}.
recently [15].

Quantum states can encode information (e.g., by amplitude, phase, or both)
that may only be extracted by analyzing the states as a whole. Stimulated by
the rapid development in quantum communication and quantum cryptography, the
question of how to determine the actual state of a quantum system has gained
renewed interest. Recently, a large number of significant schemes for
diagnosing the states of a system have been proposed based on the quantum
tomography technique (i.e., reconstruction of the states by measuring enough
observables, given many copies of a quantum state) [16]. However, for some
solid state qubit systems, it is very difficult to implement quantum
tomography since usually very few, if not just one, observables can be
measured.

In this work, we propose a method for extracting the relative phase of a $n$%
-qubit two-component entangled state with the general form of 
\begin{equation}
\sqrt{p}\left| i_1i_2...i_n\right\rangle +e^{i\varphi }\sqrt{1-p}\left| 
\overline{i}_1\overline{i}_2...\overline{i}_n\right\rangle ,
\end{equation}
(where $i_l,\overline{i}_l\in \left\{ 0,1\right\} ,$ $\overline{i}_l=1-i_l$, 
$\varphi $ is a relative phase, and $0<p<1$). The method presented here
operates essentially by local Hadamard transformations and measurements
along a single basis (the $z$ basis: $\left| i\right\rangle $ and $\left| 
\overline{i}\right\rangle $). This work is of significant interest because
it provides a way to extract the relative phase from the state (1) and
therefore can be used to determine an {\it arbitrary} multiqubit
two-component entangled state. In addition, the extraction of the relative
phase may have other applications in quantum information theory and high
precision measurements [17,18]. In the following, we will further show how
to distinguish the states (1) with arbitrary weighing factors (any
entanglement degree) and relative phases from a class of multiqubit mixed
states.

Before our presentation, let us give a brief discussion on the states (1).
For $p=1/2$ and $n=2,$ the states (1) are reduced to the well known Bell
states; while for $p=1/2$ and $n>2,$ they are called
Greenberger-Horne-Zeilinger (GHZ) states [19]. The Bell or GHZ states play
an important role in quantum information processing and communication and
are of great interest in the foundations of quantum mechanics and
measurement theory [1,20-22]. Recently, there has been much interest in
generation and engineering of the Bell or GHZ states and several methods
have been proposed for their generation in physical systems such as atoms,
ions, quantum dots, charge qubits, and flux qubits [23-31]. On the other
hand, for $0<p<1/2,$ the states (1) are partially (non-maximally) entangled
states, which may be caused by non-exact control of operations or
decoherence. As is well known, a $n$-qubit maximally entangled state can be
extracted from a set of partially entangled states with the same form of Eq.
(1) by using entanglement purification protocol [32,33]. And, quantum
communication (e.g., teleportation and cryptograph) based on partially
entangled states of Eq. (1) have been discussed recently [34-36].

Now let us see how to extract the relative phase $\varphi $ of state (1). It
can be shown that if a Hadamard transformation described by $\left|
0\right\rangle \rightarrow \frac 1{\sqrt{2}}\left( \left| 0\right\rangle
+\left| 1\right\rangle \right) $ and $\left| 1\right\rangle \rightarrow 
\frac 1{\sqrt{2}}\left( \left| 0\right\rangle -\left| 1\right\rangle \right) 
$ is performed on each qubit, then the state (1) becomes 
\begin{eqnarray}
&&\ \ \ \ 2^{-n/2}\ \left( \sqrt{p}+e^{i\varphi }\sqrt{1-p}\right)
\sum_{\{x_l\}}\left( \pm \left| \{x_l\}\right\rangle \right)  \nonumber \\
&&\ \ \ \ \ \ \ +2^{-n/2}\left( \sqrt{p}-e^{i\varphi }\sqrt{1-p}\right)
\sum_{\{y_l\}}\left( \pm \left| \{y_l\}\right\rangle \right) .
\end{eqnarray}
Here, $\left| \{x_l\}\right\rangle =\left| x_1x_2...x_n\right\rangle $ and $%
\left| \left\{ y_l\right\} \right\rangle =\left| y_1y_2...y_n\right\rangle $
are computational basis states of the $n$ qubits ($x_l,y_l\in
\{0,1\};l=1,2,...n$), and $\sum_{\{x_l\}}\left( \pm \left|
\{x_l\}\right\rangle \right) $ [$\sum_{\{y_l\}}\left( \pm \left|
\{y_l\}\right\rangle \right) $] is a sum over all possible basis states $%
\left| \{x_l\}\right\rangle $ ($\left| \{y_l\}\right\rangle $) each
containing an $even$ ($odd$) number of ``1''s. For instance, when $n=4,$ $%
\sum_{\{x_l\}}\left( \pm \left| \{x_l\}\right\rangle \right) =\pm \left|
0000\right\rangle \pm \left| 1100\right\rangle \pm \left| 1010\right\rangle
\pm \cdot \cdot \cdot \pm \left| 1111\right\rangle .$ In Eq. (2), the $\pm $
signs for each term $\left| \{x_l\}\right\rangle $ ($\left|
\{y_l\}\right\rangle $) depend on the term $\left| i_1i_2...i_n\right\rangle 
$ involved in the original state (1). That is, the choice for the $+$ sign
or the $-$ sign is determined by the number of ``1''s contained in the
binary computational basis state $\left| i_1i_2...i_n\right\rangle $ and the
arrangement of ``1''s in $\left| i_1i_2...i_n\right\rangle .$ It should be
mentioned that the choice for the $\pm $ signs does not affect the
measurement results since all measurements throughout this paper are made on
individual qubits along the $z$ basis.

Eq. (2) demonstrates that when the $n$ qubits were initially in the state
(1), if a measurement in $z$-basis is made on each qubit after a Hadamard
transformation on each qubit, the probability of finding an $even$ number of
qubits in the state $\left| 1\right\rangle $ would be given by 
\begin{equation}
p_{even}=\frac 12+\sqrt{p(1-p)}\cos \varphi .
\end{equation}

The result (3) shows that the relative phase $\varphi $ of the entangled
state\ (1) can be determined using the procedure described above. Reasons
for this are as follows. First, the value of $p$ can be obtained using a
simple measurement in $z$-basis on each qubit alone, given many copies of
the state (1). Second, the probability $p_{even}$ can be determined by
measurements in $z$-basis on each qubit after performing a Hadamard
transformation on each qubit. Therefore, one can determine the form of a
multiqubit two-component entangled state (1), with the prior knowledge that
the qubit system is prepared in a certain but unknown state belonging to a
finite set of possible states (1). Finally, it is noted that the state (1)
is reduced to a pure state of a single qubit when $n=1$. Hence, a pure state 
$\sqrt{p}\left| 0\right\rangle +e^{i\varphi }\sqrt{1-p}\left| 1\right\rangle 
$ of a single qubit with arbitrary $p$ and $\varphi $ can be confirmed using
the above procedure.

In what follows our purpose is to show how to distinguish the entangled
state (1) from a class of $n$-qubit mixed state: 
\begin{equation}
\rho _{mix}=p\left| i_1i_2...i_n\right\rangle \left\langle
i_1i_2...i_n\right| +(1-p)\left| \overline{i}_1\overline{i}_2...\overline{i}%
_n\right\rangle \left\langle \overline{i}_1\overline{i}_2...\overline{i}%
_n\right| .
\end{equation}
Note that for the state (1) the probabilities of the $n$ qubits being in $%
\left| i_1i_2...i_n\right\rangle $ and $\left| \overline{i}_1\overline{i}%
_2...\overline{i}_n\right\rangle $ are given by $p$ and $1-p$, respectively.
However, the same results are obtained when the $n$ qubits are in the mixed
state (4). Hence, one needs to distinguish the state (1) from the mixed
state (4) to make sure that the $n$ qubits are indeed prepared in the
entangled state (1).

It is straightforward to show that if a Hadamard transformation is performed
on each qubit, then the density operator (4) becomes 
\begin{equation}
\ \ \ \ \ \ \ \ \frac p{2^n}\sum_{\{x_l^{\prime }\}}\left( \pm \left|
\{x_l^{\prime }\}\right\rangle \right) \otimes \sum_{\{x_l^{\prime
}\}}\left( \pm \left\langle \{x_l^{\prime }\}\right| \right) +\frac q{2^n}%
\sum_{\{x_l^{\prime }\}}\left( \pm \left| \{y_l^{\prime }\}\right\rangle
\right) \otimes \sum_{\{x_l^{\prime }\}}\left( \pm \left\langle
\{y_l^{\prime }\}\right| \right) .
\end{equation}
Here, $\left| \{x_l^{\prime }\}\right\rangle =\left| x_1^{\prime
}x_2^{\prime }...x_n^{\prime }\right\rangle $ and $\left| \left\{
y_l^{\prime }\right\} \right\rangle =\left| y_1^{\prime }y_2^{\prime
}...y_n^{\prime }\right\rangle $ are computational basis states of the $n$
qubits ($x_l^{\prime },y_l^{\prime }\in \{0,1\};l=1,2,...n$), and $%
\sum_{\{x_l^{\prime }\}}\left( \pm \left| \{x_l^{\prime }\}\right\rangle
\right) $ [$\sum_{\{y_l^{\prime }\}}\left( \pm \left| \{y_l^{\prime
}\}\right\rangle \right) $] is a sum over all possible basis states. For
instance, when $n=3,$ $\sum_{\{x_l\}}\left( \pm \left| \{x_l^{\prime
}\}\right\rangle \right) $ or $\sum_{\{y_l^{\prime }\}}\left( \pm \left|
\{y_l^{\prime }\}\right\rangle \right) =\pm \left| 000\right\rangle \pm
\left| 001\right\rangle \pm \left| 010\right\rangle \pm \left|
100\right\rangle \pm \left| 011\right\rangle \pm \left| 101\right\rangle \pm
\left| 110\right\rangle \pm \left| 111\right\rangle .$ The $\pm $ signs for
each term $\left| \{x_l^{\prime }\}\right\rangle $ ($\left| \{y_l^{\prime
}\}\right\rangle $) are determined by the number of ``1''s and the
arrangement of ``1''s in the basis state $\left| i_1i_2...i_n\right\rangle $
involved in the density operator (4).

Eq. (5) shows clearly that for the mixed state (4), if each qubit is
measured in $z$-basis after a Hadamard transformation on each qubit, the
probability for an $even$ number of qubits being found in the state $\left|
1\right\rangle $ would be $1/2.$ On the other hand, as shown above, for the $%
n$-qubit initial state (1), if the same operation is performed on each
qubit, the probability of an $even$ number of qubits being found in the
state $\left| 1\right\rangle $ is given by Eq. (3). Hence, in general, when $%
p_{even}\neq 1/2,$ it can be concluded that the $n$ qubits are in the state
(1).

However, for the special case of $\varphi =\pm \pi /2,$ the entangled state
(1) also has $p_{even}=1/2$. Namely, the following $n$-qubit entangled state 
\begin{equation}
\sqrt{p}\left| i_1i_2...i_n\right\rangle +e^{\pm i\pi /2}\sqrt{1-p}\left| 
\overline{i}_1\overline{i}_2...\overline{i}_n\right\rangle
\end{equation}
could not be distinguished from the mixed state (4) using the procedures
described above. However, we note that for this special case, the above
method for entanglement verification based on single-basis measurement can
still work. This is because, by performing a local phase shift operation on 
{\it any} one of the $n$ qubits, described by $\left| i\right\rangle
\rightarrow \left| i\right\rangle ,\left| \overline{i}\right\rangle
\rightarrow e^{i\pi /2}\left| \overline{i}\right\rangle $, the state (6) is
transformed into 
\begin{equation}
\sqrt{p}\left| i_1i_2...i_n\right\rangle \mp \sqrt{1-p}\left| \overline{i}_1%
\overline{i}_2...\overline{i}_n\right\rangle ,
\end{equation}
i.e., the state (1) with $\varphi =\pi $ or $0.$ The state (7) can obviously
be distinguished from the mixed state (4) using the method described above
because for them one has $p_{even}=\frac 12\mp \sqrt{p(1-p)}\neq 1/2$ from
Eq. (2).

It should be noticed that the technique of distinguishing the entangled
state (1) from the mixed state (4) by local operations and measurements on
individual qubits was previously presented in the generation of entanglement
of polarized photons (for details, see references [37-39]). However, we
point out that in this work we considered a general case, i.e., the $n$%
-qubit entangled state of Eq. (1) with arbitrary $p$ and $\varphi $ ($%
0<p<1). $ Namely, the state (1) considered here could be maximally-entangled
states with $p=1/2$ and an arbitrary relative phase $\varphi ,$ or
partially-entangled states with $0<p<1/2$ and an arbitrary relative phase $%
\varphi .$ The results presented above are valid for any qubit system. Our
work regarding the distinction between the generic state (1) and the mixed
state (4) is a generalization of the previous work [37-39] to a {\it %
multiqubit} two-component entangled state with a {\it general form, }which
is of great interest in both theory and experiment by itself.

A single-qubit Hadamard transformation can be easily achieved for many
physical qubit systems. For examples, it can be done by locally rotating the
polarization of a photon, by applying a $\pi /2$ microwave pulse resonant
with the transition between the two lowest levels (the two logical states of
a qubit) $\left| 0\right\rangle $ and $\left| 1\right\rangle $ of a SQUID
(superconducting quantum interference device) [31], or by applying a $\pi /2$
two-photon Raman resonance pulse to the two lowest levels of an atom with a $%
\Lambda $-type three-level structure.

As a matter of fact, based on $\left| 0\right\rangle =\left| +\right\rangle
+\left| -\right\rangle $ and $\left| 1\right\rangle =\left| +\right\rangle
-\left| -\right\rangle ,$ it is noted that Hadamard transformations are not
necessary for either extracting the relative phase $\varphi $ or verifying
the entangled state (1), because the same results can be obtained by
measuring each qubit in the $x$ basis $\left| +\right\rangle $ and $\left|
-\right\rangle ,$ instead of a Hadamard transformation followed by a
measurement in the $z$ basis $\left| 0\right\rangle $ and $\left|
1\right\rangle .$ However, it should be mentioned that the single-qubit
Hadamard transformations are needed for some important physical qubit
systems, for which usually only one natural measurement basis ( the $z$
basis) is available. For instance, for superconducting charge (flux) qubits,
which are promising candidate for scalable quantum information processing,
it is very difficult to perform measurement in basis other than the charge
(flux) degree of freedom.

In summary, we have shown that an arbitrary relative phase can be extracted
from a multiqubit two-component entangled state based on local single-qubit
Hadamard transformations and measurements along a single basis only. The
method is relatively easy to be realized because there is no need for
two-qubit operations and the qubits are allowed to be well separated in
space. The present work provides a way to determine an arbitrary multiqubit
two-component entangled state through the extraction of the relative phase,
which is of great importance to quantum communication in general and quantum
key distributions in particular. In addition, extracting the phase may have
other applications in quantum information theory and high precision
measurements. Finally, a detailed discussion on distinguishing the entangled
state (1) (with an arbitrary entanglement degree and relative phase) from a
class of multiqubit mixed states has been presented.

This work was partially supported by National Science Foundation
(DMR-0325551), and by AFOSR, NSA, and ARDA through DURINT grant
(F49620-01-1-0439).

\end{document}